\documentclass[pre,showpacs,groupedaddress]{revtex4}

\usepackage[latin1]{inputenc}
\usepackage{bm}
\usepackage[hypertex]{hyperref}
\usepackage{txfonts}

\newcommand{\average}[1]{\left\langle{#1}\right\rangle}

\newcommand{\D}{\mathrm{d}}
\newcommand{\E}{\mathrm{e}}
\newcommand{\I}{\mathrm{i}}
\newcommand{\DD}{\mathcal{D}}
\newcommand{\II}{\mathcal{I}}
\newcommand{\QQ}{{\mathcal{Q}}}
\newcommand{\Liouville}{\mathcal{L}}
\newcommand{\de}[2]{\frac{\partial #1}{\partial #2}}

\renewcommand{\vec}[1]{\bm{#1}}
\newcommand{\bG}{\bm{G}}
\newcommand{\br}{{\vec{r}}}
\newcommand{\bp}{{\vec{p}}}
\newcommand{\bof}{{\vec{f}}}
\newcommand{\be}{{\vec{\eta}}}

\newcommand{\bg}{{\vec{\gamma}}}

\begin{document}
\title{Fluctuation relations for a driven Brownian particle}
\author{A. Imparato}\thanks{Present address: Dipartimento di Fisica, Politecnico di Torino,
C.~Duca degli Abruzzi 24, 10129 Torino (Italy).}
\author{L. Peliti}
\affiliation{Dipartimento di Scienze Fisiche,
INFN-Sezione di Napoli, CNISM-Sezione di Napoli \\
Universit\`a ``Federico II'', Complesso Monte S. Angelo, I--80126
Napoli (Italy)} 

\begin{abstract}
    We consider a driven Brownian particle, subject to both
    conservative and non-conservative applied forces, whose
    probability evolves according to the Kramers equation.
    We derive a general fluctuation relation, expressing the ratio
    of the probability of a given Brownian path in phase
    space with that of the time-reversed path, in terms
    of the entropy flux to the heat reservoir. This fluctuation
    relation implies those of Seifert, Jarzynski and
    Gallavotti-Cohen in different special cases.
\end{abstract}

\maketitle

\section{Introduction}
Fluctuation relations of increasing generality have been derived
in the recent years, starting from the Evans-Searles
relation~\cite{Evans} and the Gallavotti-Cohen theorem~\cite{GC},
via the Jarzynski equality~\cite{Jarzynski}, the Hatano-Sasa
relation~\cite{HS}, to arrive at the recent and rather general
Seifert relation~\cite{Seifert}, from which the previous
identities can be derived by simple manipulations. These results
have been obtained by considering situations of increasing
generality, and it appears that a sufficient condition for
fluctuation relations to hold is the existence of
\textit{mechanical} equilibrium at each instant of
time~\cite{Astumian}. However, this condition is not necessary, as
shown by the derivation by Kurchan~\cite{Kurchan} of some
fluctuation relations for a Brownian particle with inertia.

We wish to point out that it is possible to obtain a compact
derivation of the fluctuation relations by deriving a key relation
between the probability of a path in \textit{phase space} and that
of its time-reversed image, for a particle with inertia. This
relation encompasses the relation derived by Crooks~\cite{Crooks}
for a particle subject to conservative forces and extended by
Seifert~\cite{Seifert} to arbitrary forces in the overdamped
regime. Our derivation is inspired by Kurchan's
one~\cite{Kurchan}, but is made more compact and more general by
considering the generating functional of the conditional
probabilities for the paths.

As a bonus, we obtain an explicit expression for the entropy
production and the heat flow for a Brownian particle with inertia,
which generalizes the expression introduced by
Sekimoto~\cite{Sekimoto}, valid in the overdamped regime.

The key relation between the probability of a path $\omega$,
conditioned on its initial point, and that of the reverse path
$\tilde\omega$, conditioned on \textit{its} initial point, is
derived in sec.~\ref{identity:sec}. The relation involves a
functional of the path, whose physical interpretation as the
entropy flow into the heat bath is given in sec.~\ref{heat:sec}.
In sec.~\ref{OM:sec}, using the formalism previously introduced, the explicit expression of path probability as a form of the
Onsager-Machlup functional is recovered, and the fluctuation relation is shown to hold for such a functional.
A brief discussion follows, while the derivation of the relation between
the time derivatives of ``global'' and ``local'' entropies is
given in the Appendix.

\section{A generalized Crooks identity}\label{identity:sec}
We consider a particle in $d$ dimensions, whose evolution is
described by the Langevin equation
\begin{equation}
m  \ddot{\vec{r} }= -\zeta \dot{\vec{ r}}  - \de
{U_\mu(\vec{r})}{\vec{ r}}+ \vec{f}_\mu +\be(t), \label{newt}
\end{equation}
where $\mu$ is a time-dependent parameter, and $\vec{f}_\mu$ is a
non conservative force. We assume that $\be(t)$ is a
delta-correlated white noise with variance $2 \zeta T$.
Boltzmann's constant is set equal to 1 throughout.

We denote by $x=(\bp, \br)$ the microstate, and by $P(x,t)$ the
probability distribution function (pdf) of the
process~(\ref{newt}) in phase space. This function evolves
according to the differential equation
\begin{equation}
\de{P(x,t)} t= \Liouville_{\mu(t)} P(x,t), \label{kram}
\end{equation}
where $\Liouville_{\mu}$ is the Kramers operator
\begin{eqnarray}
    \Liouville_\mu \,P
    &=&\frac{\partial}{\partial\br }\cdot\left(-\frac{\bp}{m}
    \,P\right)+\frac{\partial}{\partial\bp}\cdot
    \left[\left(\zeta \frac{\bp}{m}
      +\frac{\partial U_\mu}{\partial \br}
      -\bof_\mu\right)P+\zeta T \frac{\partial P}{\partial\bp}
     \right].
\label{op:lio}
\end{eqnarray}
Let us define the time-reversal operator $\II$ by
\begin{equation}
    \II (\vec p,\vec r)=(-\vec p,\vec r),
\end{equation}
and the associated operator $\QQ_t$ by
\begin{equation}
    \QQ_t \,\psi(x)=\E^{-E(x,t)/T}\,\psi(\II x).
\label{defQ}
\end{equation}
We assume that the system energy, defined by
\begin{equation}
E(x,t)=\sum_{i=1}^d \frac{p_i^2}{ 2 m} +U_{\mu(t)}(\br),
\end{equation}
is invariant under time inversion:
\begin{equation}
E(\II x,t)=E(x,t).
\end{equation}
The microscopic reversibility of the process (\ref{newt}) implies
the following relation
\begin{equation}
\QQ_t^{-1} \Liouville_{\mu(t)} \QQ_t=\Liouville^\dag_{\mu(t)}
    +\frac{\bof_{\mu(t)}}{T}\cdot\frac{\bp}{m}.
\end{equation}

In the following the symbol $\omega$ will indicate
an arbitrary path in the system phase space:
\begin{equation}\label{omega:eq}
    \omega:\quad t\stackrel{\omega}{\longrightarrow} x(t).
\end{equation}
For any arbitrary function $a(x,t)$, and any arbitrary path
$\omega$ in the system phase space we define the quantity
$A(t,t_0,\omega)$ as the integral of $a(x,t)$ over the interval
$\left[t_0,t\right]$ along the given  path $\omega$:
\begin{equation}\label{Aint:eq}
A(t,t_0, \omega)\equiv\int ^t_{t_0} \D t'\; a(x(t'),t').
\end{equation}
The joint probability distribution function $\Phi(x,A,t)$
evolves according to the differential equation
\begin{equation}
\de \Phi t =\Liouville_{\mu(t)}\Phi -a(x,t) \,\de \Phi A.
\end{equation}

Let us define the function
\begin{eqnarray}
\Psi(x,t;x_0,t_0;\mu,a)&=&\int^{x(t)=x}_{x(t_0)=x_0} \DD \omega\; \mathcal{P}(\omega)
\int \D A\; \delta(A-A(t,t_0,\omega) )\,\E^{A(t,t_0,\omega)}\nonumber\\
&=& \int \D A\; \E^{A}\, \Phi(x,A,t), \label{defPsi0}
\end{eqnarray}
where we make explicit the fact that $\Psi$ depends on the initial
condition $x_0,t_0$, and is a functional of $\mu(t)$ and $a(x,t)$.
Then $\Psi$ satisfies the following equation:
\begin{equation}\label{psiev:eq}
\de \Psi t =\Liouville_{\mu(t)} \Psi +a(x,t)\,\Psi.
\end{equation}

A number of fluctuation relations can be easily derived from this expression.
To illustrate this
point, we report here the derivation of the Hatano-Sasa relation~\cite{HS}.
Let
\begin{equation}
P^{\mathrm{SS}}(x,\mu)= \E^{-\phi(x,\mu)}
\end{equation}
be the steady-state distribution function associated to the
operator $\Liouville_{\mu}$, which satisfies
\begin{equation}
\Liouville_{\mu}P^{\mathrm{SS}}(x,\mu)=0.
\end{equation}
Let us  choose the arbitrary function $a(x,t)$ as follows:
\begin{equation}
    a^*(x,t)= \dot \mu(t)\de{\ln
    P^{\mathrm{SS}}}{\mu}=-\dot\mu(t)\,\de{\phi}{\mu},
\end{equation}
so that eq.~(\ref{Aint:eq}) reads
\begin{equation}
    A=-\int^t_{t_0} \D t' \dot \mu(t') \left.\de{\phi}{\mu}\right|_{x(t'),\mu(t')}.
\end{equation}
Using the definition of $\Psi$, eq.~(\ref{defPsi0}), one finds
\begin{eqnarray}
    \average{\delta(x-x(t))\,\E^{A(t)}}&=&\int \D A\;
    \E^{A}\,\Phi(x,A,t)
    =\Psi(x,t;x_0,t_0;\mu,a^*). \label{hs1}
\end{eqnarray}
But $\Psi$ is solution of eq.~(\ref{psiev:eq}), which
for the particular choice $a=a^*$ takes the form
\begin{equation}
    \de \Psi t=\Liouville_{\mu(t)}\Psi-\dot \mu \de{\phi}{\mu}\, \Psi.
\end{equation}
The solution of this last equation, satisfying the initial
condition $\Psi(x,t_0)=P^\mathrm{SS}(x,\mu(t_0))$, is
$\Psi=\E^{-\phi(x,\mu(t))} =P^{\mathrm{SS}}(x,\mu(t))$, and thus
eq.~(\ref{hs1}) becomes
\begin{equation}
\average{\delta(x-x(t))\,\E^{A(t)}}= P^{\mathrm{SS}}(x,\mu(t)),
\end{equation}
which is the Hatano-Sasa identity, and reduces to the Jarzynski
equality when the non-conservative force $\vec f_\mu$ vanishes
throughout.

Now, let the functional $ \widetilde\Psi(x,t;x_0,t_0;\mu,a)$ be defined
by
\begin{equation}
\widetilde \Psi(x,t;x_0,t_0;\mu,a)\equiv \QQ^{-1}_t
\Psi(x,t;x_0,t_0;\mu,a)\, \QQ_{t_0}.
\end{equation}
One can easily check that $\widetilde\Psi$ satisfies the evolution
equation
\begin{equation}
    \frac{\partial \widetilde\Psi}{\partial t}={\Liouville}_t^\dag\,\widetilde\Psi
    +\left[\frac 1 T \frac{\partial E}{\partial t}+\frac{\bof_{\mu(t)}}{T}\cdot \frac{
    \bp}{m}+a(\II x,t)\right]\widetilde\Psi.
\label{eqPsitil}
\end{equation}
Using equations (\ref{defQ}) and (\ref{defPsi0}), we obtain the
following equality
\begin{equation}
    \widetilde\Psi(\II x,t;\II x_0,t_0;\mu,a)
    =\E^{[E(x,t)-E(x_0,t_0)]/T}\Psi(x,t;x_0,t_0;\mu,a).
\label{Psiinv1}
\end{equation}
Furthermore, upon integration of eq.~(\ref{eqPsitil}), we obtain
\begin{equation}
    \widetilde\Psi(\II x,t;\II x_0,t_0;\mu,a)=\Psi(\II x_0,t;\II x,t_0;\tilde\mu,\tilde a+w),
\label{Psiinv2}
\end{equation}
where we have defined, $\forall t'\in[t_0,t]$,
\begin{eqnarray}
    \tilde t'&=& t-(t'-t_0);\\
    \tilde \mu(t')&=&\mu(\tilde t');\\
    \tilde a(x,t')&=&a(\II x,\tilde t');\\
    w(x,t)&=&\frac 1 T \left[\frac{\partial E}{\partial
    t}+\bof_{\mu(t)}    \cdot\frac{\bp}{m}\right].
\end{eqnarray}
Thus, substituting eq.~(\ref{Psiinv1}) into eq.~(\ref{Psiinv2}) we obtain
\begin{eqnarray}\label{invtemp:eq}
    \Psi(x,t;x_0,t_0;\mu,a)&=&\Psi(\II x_0,t;\II x,t_0;\tilde\mu,\tilde a+w)\,\E^{-[E(x,t)-E(x_0,t_0)]/T}.
\end{eqnarray}

We choose $a(x,t)$ as
\begin{equation}
a(x,t)=\lambda(t) x=\vec \lambda^{(p)}(t)\cdot \vec p+\vec
\lambda^{(r)}(t)\cdot \vec r.
\end{equation}
Then, using the definition of $\Psi$, eq.~(\ref{defPsi0}), we
obtain the generating functional for the path probabilities
\begin{eqnarray}
   \Psi(x,t;x_0,t_0;\mu,\lambda\,x)&=&\int_{x(t_0)=x_0}^{x(t)=x} \DD \omega\;
    \mathcal{P}(\omega)\,\exp\left(\int_{t_0}^t\D t'\; \lambda(t')
    x(t')\right).
\label{defPsi}
\end{eqnarray}
Now, for each path $\omega$, starting from $x_0$ and finishing in
$x$, the following equality holds
\begin{equation}\label{work:eq}
    E(x,t)-E(x_0,t_0)=\int_{t_0}^t\left(\frac{\partial E}{\partial
    t'}\,\D t'
    +\left.\frac{\partial E}{\partial x}\right|_{x(t')}\,\D x(t')\right).
\end{equation}
In this expression, the integral has to be interpreted according
to Stratonovich, otherwise additional terms would appear. The
relation (\ref{defPsi}) holds for arbitrary $\lambda$, and thus
leads to an analogous relation for the probabilities of the paths
$\omega$ conditional on their initial points. Indeed, given the
path $\omega$ defined by (\ref{omega:eq}), we formally have
\begin{equation}\label{pathprob:eq}
    \mathcal{P}(\omega,\mu|x_0,t_0)=\int \DD\lambda
    \;\E^{-\int_{t_0}^t\D t'\;\lambda(t')
    x(t')}\,\Psi(x,t;x_0,t_0;\mu,\lambda\,x),
\end{equation}
where we have introduced the functional integration over $\lambda$
from $-\I\infty$ to $+\I\infty$. Applying this operation to
eqs.~(\ref{invtemp:eq}) and (\ref{defPsi}) we obtain the central
result
\begin{equation}\label{central:eq}
    \mathcal{P}(\omega,\mu|x_0,t_0)=\mathcal{P}(\widetilde\omega,\tilde\mu|\II x,t_0)\,
    \exp\left[\int^x_{x_0} \frac{\D Q_2} T\right],\label{inv:eq}
\end{equation}
where
\begin{equation}
\D Q_2(t)=-\left.\frac{\partial E}{\partial x}\right|_{x(t)} \,\D x(t)
    +\bof_{\mu(t)}\cdot \frac{\bp(t)}{m}\,\D t.
\label{dq2:eq}
\end{equation}
Thus Crooks's and Seifert's identities~(\cite[eq.~(13)]{Crooks},
\cite[eq.~(14)]{Seifert}) hold also in the case in which the
Brownian particle is subject to conservative and nonconservative
forces, which do not necessarily equilibrate mechanically with the
friction force at each instant in time. Note that the expression
of $\D Q_2$ must be supplemented by its interpretation as a
Stratonovich differential.

Seifert~\cite{Seifert} has taught us how to derive the relevant
fluctuation relations from the identity (\ref{central:eq}), by
introducing carefully chosen ``initial'' and ``final'' pdf's
$p_0(x_0)$, $p_1(\II x)$. For example, let $p_0(x_0)$ be an
arbitrary normalized initial condition at time $t_0$, and $p_1(x)$
be the corresponding evolved pdf at time $t$, and define
\begin{equation}
  R=\int_{t_0}^t \frac{\D Q_2}{T}+\ln \frac{p_0(x_0)}{p_1(x)}.
\end{equation}
Then
\begin{eqnarray}
\average{\delta(x-x(t))\,\E^{-R}}&=&\int \D x_0 \int \DD \omega\;
\mathcal{P}(\widetilde\omega,\tilde\mu|\II x, t_0) \,p_1(x) =
p_1(x).
\end{eqnarray}
This result can be obtained by changing the integral over $\omega$
into an integral over $\widetilde\omega$ and by exploiting the
normalization of the conditional probability of the evolution from
$\II x$ to $\II x_0$ over the final state $\II x_0$. Then one can
interpret $R$ as the total entropy change, being the sum between
the entropy given to the heat bath and the entropy change of the
particle, as we shall see in the next section. The Jarzynski
equality also follows from eq.~(\ref{central:eq}) by taking for
$p_0(x_0)$ and $p_1(x)$ the equilibrium distributions
corresponding to the initial and final states. If one takes
instead $\mu$ as a constant, and collects together the paths
$\omega$ which yield a given value $\Sigma$ of $\int\D Q_2/T$, one
obtains a ``transient'' form of the Gallavotti-Cohen relation:
\begin{equation}\label{GC1:eq}
    \frac{P(\Sigma)}{P(-\Sigma)}=\E^\Sigma.
\end{equation}
The Gallavotti-Cohen relation follows in the limit of long times,
by setting
\begin{equation}\label{GC}
    P(\Sigma)\simeq \E^{t \Lambda(\sigma)},
\end{equation}
where
\begin{equation}
    \Sigma= t \,\sigma,
\end{equation}
and where $\Lambda(\sigma)$ (implicitly defined by eq.~(\ref{GC})
itself) is the large-deviation function for the entropy production
rate.

\section{Entropy production and heat flow}\label{heat:sec}
We provide here the physical interpretation of the quantity $\D
Q_2(t)$. Let us define the local entropy
\begin{equation}
s(x,t)=-\ln P(x,t).
\end{equation}
The Shannon entropy of the system is thus given by
\begin{equation}
S(t)\equiv \average {s(x,t)}=-\int \D x\; P(x,t)\, \ln P(x,t).
\end{equation}
The time derivative of this quantity is given by
\begin{eqnarray}
\de {S(t)}{ t}&=&-\int \D x\; \left\{\de {P(x,t)}{t}
\, \ln\left[P(x,t)\right]+ P(x,t) \de {\ln P(x,t)} t\right\}\label{dS0}\\
&=&-\int \D x\; \de {P(x,t)}{t} \,\ln P(x,t)
\label{dS1}\\
&=&\int \D x\; \left[\zeta T \left(\frac {\bp}{T m} + \de {\ln
P}{\bp}\right)^2 P - \zeta  \frac {\bp} m \cdot\left(\frac {\bp}
{Tm} + \de {\ln P}{\bp}\right) P\right]\label{dS2},
\end{eqnarray}
where the integral of the second term of eq.~(\ref{dS0}) vanishes
because of the conservation of the pdf normalization. For the
details of the calculations see the appendix,
equation~(\ref{appdS1}). On the other hand we have
\begin{eqnarray}
\average{\left.\frac{\D s(x(t),t)}{\D
t}\right|_{x(t)=x}}_{\be}&=&\int \D x\; P(x,t) \left[\zeta T
\left(\frac {\bp}{T m}+ \de {\ln P}{\bp}\right)^2 -\zeta  \frac
{\bp}m\cdot \left(\frac {\bp} {Tm} + \de {\ln P}{\bp}\right)
\right]. \label{ds1}
\end{eqnarray}
where $\average {\cdots}_{\be}$ indicates the average over the pdf
$P(x,t)$ and over the realizations of the process. For the details
of the calculations see the appendix,
equations~(\ref{appds0}--\ref{appds2}). Thus we obtain the result
\begin{equation}
\frac{\D S(t)}{\D t}=\average{\left.\frac{\D s(x(t),t)}{\D
t}\right|_{x(t)=x}}_{\be}. \label{eqSs}
\end{equation}

Let us now define the quantity
\begin{eqnarray}
\D Q_\mathrm{p}&=&\D Q_2 +T \left(\de{s}{\br} \cdot\D \br
    +\de{s}{\bp}\cdot \D \bp\right)\\
    &=&-\left.\frac{\partial E}{\partial x}\right|_{x(t)} \,\D x(t)
    +\bof_{\mu(t)}\cdot \frac{\bp(t)}{m}\,\D t
    +T \left(\de{s}{\br}\cdot \D \br +\de{s}{\bp}\cdot \D
    \bp\right).\nonumber
\label{defqp}
\end{eqnarray}
We have
\begin{eqnarray}
\average{\left.\frac{\D Q_\mathrm{p}(x(t),t)}{\D
t}\right|_{x(t)=x}}_{\be}&=&
\average{\zeta \left(\frac{\bp}{ m }+T \de {\ln P}{\bp}\right)^2}\\
&=&T \frac{\D S(x,t)}{\D t}+ \zeta \frac{\bp}{ m }\cdot\left(\frac
{\bp}{ m }+T \de {\ln P}{\bp}\right).\nonumber
\end{eqnarray}
The last equation indicates that the average of $\dot
Q_\mathrm{p}/T$ is a non negative quantity, and can thus be
interpreted as the entropy production rate of the process.
Furthermore, rearranging eq.~(\ref{defqp}) we have
\begin{equation}
\D Q_\mathrm{p}-\D Q_2=T  \left(\de{s}{\br}\cdot \D \br
+\de{s}{\bp}\cdot \D \bp\right).
\end{equation}
Since one finds
\begin{equation}
\average{\left.\frac{\D s(x(t),t)}{\D
t}\right|_{x(t)=x}}_{\be}=\average{\de{s}{\br} \cdot\D \br
+\de{s}{\bp} \cdot\D \bp}_{\be},
\end{equation}
(see appendix,  eqs.~(\ref{appds0})-(\ref{appds1})), exploiting
eq.~(\ref{eqSs}), one ends up with
\begin{equation}
\Delta S\equiv\int^t_{t_0}\frac{\D S(t)}{\D
t}=\frac{1}{T}\int^t_{t_0} \average{\D Q_\mathrm{p}-\D Q_2}_{\be}.
\end{equation}
Thus, we can interpret the quantity $\D Q_2/T$ as the entropy flow
into the heat reservoir.

\section{Fluctuation relations and the Onsager--Machlup functional}\label{OM:sec}
 Following the approach outlined in
\cite{noi}, and introducing the variable $\bg^{(p)}$ conjugate to
$\bp$ and $\bg^{(r)}$ conjugate to $\br$, we can express the
functional $\Psi$ as
\begin{eqnarray}
    \Psi(x,t;x_0,t_0;\mu,a)
    =\lim_{\mathcal{N}\to\infty}
    \int\prod_{k=1}^\mathcal{N}
    \frac{\D \gamma_k\,\D x_k}{(2\pi\I)^{2d}}\;
    \delta(x-x_\mathcal{N})
    \exp\left\{\sum_{k=1}^\mathcal{N}
    \left[\gamma_k(x_k-x_{k-1})
    +\left(L(\gamma_k,x_{k-1},t_{k-1})+a(x_{k-1},t_{k-1})\right)\Delta t\right]\right\},
\label{psi_dis}
\end{eqnarray}
where
\begin{eqnarray}
    \Delta t&=&\frac{t-t_0}{\mathcal{N}};\\
    t_k&=&t_0+k\,\Delta t;\\
    \gamma x&=&\bg^{(p)}\cdot\bp+\bg^{(r)}\cdot\br;\\
    L(\gamma,x,t)&=&    \bg^{(r)}\cdot\left(-\frac{\bp}{m}\right)
    +\bg^{(p)}\cdot\left[\zeta\frac{\bp}{m}
      +\frac{\partial U_{\mu(t)}}{\partial\br}
        -\bof_{\mu(t)}+\zeta T\,\bg^{(p)}\right].\label{defL}
\end{eqnarray}
In the continuum limit, equation (\ref{psi_dis}) becomes
\begin{equation}
    \Psi(x,t;x_0,t_0;\mu,a)=\int_{x(t_0)=x_0}^{x(t)=x} \DD\gamma\,\DD \omega\;
    \E^{\mathcal{S}[\gamma,\omega,a]},
\end{equation}
where
\begin{equation}
   \mathcal{S}[\gamma,\omega,a]=\int_{t_0}^t\D t'\;\left[\gamma(t')\,
    \dot{x}(t')+L(\gamma(t'),x(t'),t')
      +a(x(t'),t')\right].\nonumber
\end{equation}
is the ``action" associated to the given path $\omega$.
From eq.~(\ref{inv:eq}), taking into account eq.~(\ref{dq2:eq}), we obtain
\begin{equation}
    \Psi(\II x_0,t;\II x,t_0;\tilde\mu,\tilde a+w)
    =\int_{x(t_0)=\II x}^{x(t)=\II x_0} \DD\gamma\,\DD \tilde\omega\;
    \E^{\mathcal{S}[\gamma,\tilde\omega,\tilde a+w]}=
    \int_{x(t_0)= x_0}^{x(t)=x} \DD\gamma\,\DD\omega\;
    \E^{\widetilde\mathcal{S}[\gamma,\omega,a]},
\end{equation}
where
\begin{equation}
  \widetilde\mathcal{S}[\gamma,\omega,a]=\mathcal{S}[\gamma,\omega,a] +\frac 1 T \int_{t_0}^t\D t'\;
   \left(\left.\frac{\partial E}{\partial x}\right|_{x(t')} \,\dot x(t')
    -\bof_{\mu(t')}\cdot \frac{\bp(t')}{m}\right).
\end{equation}
The second term on the rhs of the last equation corresponds to
$-\int_{t_0}^t \D Q_2$, where $\D Q_2$ is defined by
eq.~(\ref{dq2:eq}), along the given path $\omega$.

We can obtain an explicit expression for the probabilities of the
path by performing the gaussian functional integral over $\gamma$.
By taking $a(x,t)=0$ from eq.~(\ref{defPsi0})  and eq.~(\ref{psi_dis}),
we obtain that the probability of a given path $\omega$ is given by
\begin{equation}
 \mathcal{P}[\omega]=\lim_{\mathcal{N}\to\infty}
    \int\prod_{k=1}^\mathcal{N}\left[
    \frac{\D \gamma_k}{(2\pi\I)^{2d}}
    \, \E^{\gamma_k(x_k-x_{k-1})+L(\gamma_k,x_{k-1},t_{k-1})\,\Delta t}\right].
\label{pofs}
\end{equation}
If in last equation we  substitute the expression for
$L$ as given by eq.~(\ref{defL}), we obtain
\begin{eqnarray}
 \mathcal{P}[\omega]&=&\lim_{\mathcal{N}\to\infty}
\int  \prod_{k=1}^\mathcal{N}\left[\frac{\D
\bg^{(r)}_k}{(2\pi\I)^d}\frac{\D \bg^{(p)}_k}{(2\pi\I)^d}\right]\;
\exp\left\{\sum_{k=1}^\mathcal{N}\left[\bg^{(r)}_k \left(\Delta \br_k-\frac{\bp_k}{m}\,\Delta t\right)
+\Delta t\,\zeta T \left(\bg^{(p)}_k\right)^2 \right.\right.\nonumber\\
&&\qquad\left.\left.{}+\bg^{(p)}\left[\Delta \bp_k+\Delta
t\,\left(\zeta \frac {\bp_k}{m}
+\left.\de{U_{\mu(t_k)}}{\br}\right|_{\br_k}-\bof_{\mu(t_k)}\right)\right]\right] \right\}\nonumber\\
&\propto&\lim_{\mathcal{N}\to\infty}
    \prod_{k=1}^\mathcal{N} \delta\left(\Delta \br_k-\frac{\bp_k}{m}\,\Delta t\right)\,
\exp\left\{-\frac{1}{4 \zeta T\,\Delta t} \left[\Delta \bp_k+\Delta t\, \bG(\bp_k, \br_k, t_k) \right]^2 \right\}.
\label{pos1}
\end{eqnarray}
where
\begin{equation}
 \bG(\bp_k, \br_k, t_k)= \zeta \frac {\bp_k}{m}
+\left.\de{U_{\mu(t_k)}}{\br}\right|_{\br_k}-\bof_{\mu(t_k)}.
\end{equation}
Equation (\ref{pos1}) indicates that the probability of a given path
is non-zero only if the path satisfies
\begin{equation}
\bp= m \dot \br, \label{cond_r}
\end{equation}
 as expected.
Some care has to be taken in passage to the continuum limit in eq.~(\ref{pos1}):
the argument of the exponential in this equation reads
\begin{equation}
\sum_{k=1}^\mathcal{N} -\frac{\Delta t}{4 \zeta T}
\left[\frac{\Delta \bp_k}{\Delta t}+ \bG(\bp_k,\br_k,t_k) \right]^2.
\end{equation}
We consider separately the large $\mathcal{N}$ limit of  each of
the terms appearing in this sum: the first and the second term
 read
\begin{eqnarray}
    &&\lim_{\mathcal{N}\to\infty} \left(-\frac{\Delta t}{4 \zeta
    T}\right)\sum_{k=1}^\mathcal{N} \left(\frac{\Delta \bp_k}{\Delta
        t}\right)^2= -\frac{1}{4 \zeta T}\int_{t_0}^t \D t' \left(\dot
    \bp(t') \right)^2,\label{firstterm:eq}\\
    &&\lim_{\mathcal{N}\to\infty}\left( -\frac{\Delta
    t}{4 \zeta T}\right)\sum_{k=1}^\mathcal{N}  \left[
    \bG(\bp_k,\br_k,t_k) \right]^2=-\frac{1}{4 \zeta T}
    \int_{t_0}^t \D t' \bG^2(\bp,\br,t'),\label{secondterm:eq}
\end{eqnarray}
respectively.
 The expression appearing on the rhs of
eq.~(\ref{firstterm:eq}) is of course only formal, since the
weight of the functional integral is concentrated on functions
that are continuous but not differentiable. The double product reads
\begin{equation}
\lim_{\mathcal{N}\to\infty}\left( -\frac{1}{2 \zeta
T}\right)\sum_{k=1}^\mathcal{N} \left[ (\bp_k -\bp_{k-1})\cdot
\bG(\bp_k,\br_k,t_k) \right]\equiv-\frac{1}{2 \zeta T}
(I)\int \D \bp\cdot \bG(\bp,\br,t),\label{Iadef}
\end{equation}
and, as discussed in ref.~\cite{HR},  the value of the integral
depends on the discretization scheme: we have consequently used
the prefix $(I)$ indicating that the integral is the continuum
limit of an It\=o sum. We can express this quantity in a more convenient way by writing, in analogy with  It\=o's formula (see eqs. (17)-(19) in ref.~\cite{HR} and references therein)
\begin{equation}
 (I)\int \D \bp\cdot
\bG(\bp,\br,t)=\int_{t_0}^t \bG(\bp(t'),\br(t'),t')\cdot\dot \bp(t')
\D t'-\zeta T \int_{t_0}^t \D t' \left. \sum_{\alpha=1}^d \frac{\partial
G_\alpha}{\partial p_\alpha} \right|_{\bp(t'),\br(t'),t'}. \label{Itoexpr}
\end{equation}
Thus, substituting eqs.~(\ref{firstterm:eq}),
(\ref{secondterm:eq}) and (\ref{Itoexpr}), into eq.~(\ref{pos1}),
in the continuum limit eq.~(\ref{pofs}) becomes
\begin{eqnarray}
\mathcal{P}(\omega)&\propto&\exp\left[-\frac{1}{4 \zeta T}
\int^t_{t_0}\left(m \ddot \br(t')+\zeta \dot \br
+\left.\de{U_{\mu(t')}}{\br}\right|_{\br(t')}-\bof_{\mu(t')}\right)^2\D t'
+\frac {3 \zeta} {2 m} (t-t_0) \right]. \label{pOM}
\end{eqnarray}
These expressions correspond to those obtained by Onsager and
Machlup~\cite{OM} for the harmonically bound Brownian particle.
Note that the  linear term appearing in the exponential of
eq.~(\ref{pOM}), and which arises from eq.~(\ref{Itoexpr}), can be
``hidden" in the normalization of the probability function
$\mathcal{P}(\omega)$ as long as one deals with a force $\bG$
linear in $\vec{p}$.

From eq.~(\ref{pOM}), we are able to recover
eq.~(\ref{central:eq}) directly. In fact, for the inverse path
$\tilde \omega$, eq.~(\ref{pOM}) reads
\begin{eqnarray}
\mathcal{P}(\tilde \omega)&\propto&\exp\left[-\frac{1}{4 \zeta T}
\int^t_{t_0}\left(m\ddot \br(t')-\zeta \dot \br(t')
+\left.\de{U_{\mu(t')}}{\br}\right|_{\br(t')}-\bof_{\mu(t')}\right)^2
\D t' +\frac {3 \zeta} {2 m}\, (t-t_0) \right], \label{pOMin}
\end{eqnarray}
and thus we have, after straightforward manipulations,
\begin{eqnarray}
\frac{\mathcal{P}(\omega)} {\mathcal{P}(\tilde \omega)}&=&
\exp\left\{\frac 1 T \int^t_{t_0} \D t'\;\left[- \frac{\bp(t')}
m\cdot\left(\dot \bp(t') +
\left.\de{U_{\mu(t)}}{\br}\right|_{\br(t')}
 -\bof_{\mu(t')}\right)\right]\right\}  \nonumber\\
&=&\int^x_{x_0} \frac{\D Q_2}{T},
\label{OMQ2}
\end{eqnarray}
where $\D Q_2$ is given by eq.~(\ref{dq2:eq}).

While eq.~(\ref{pOM}) is well known, we wish to emphasize the
relation of the entropy flow along a given trajectory $\omega$,
with the generalized Onsager-Machlup functional for that
trajectory, expressed by eq.~(\ref{OMQ2}).

\section{Discussion}
We have obtained a compact derivation of the relation between the
probability of a given path in phase space and that of its
time-reversed image in the specular manipulation protocol. The
ratio of the probabilities has been interpreted as the entropy
flow into the heat bath, by a suitable generalization of the
expression of the heat flow. Different fluctuation relations can
be derived from this key equality by simple manipulations. The
approach can be easily applied to more complex systems, and to
situations in which the energy function is not invariant under
time reversal, due, e.g., to the presence of magnetic fields. We
have also shown how the basic relation directly follows from the
expression of the path probability first established by Onsager
and~Machlup~\cite{OM} for the harmonically bound particle. We hope
that the present work can be useful in the investigation of more
and more general aspects of the nonequilibrium thermodynamics of
small systems.

\begin{acknowledgments}
This research was partially supported by MIUR-PRIN 2004. We are
grateful to G. E. Crooks for suggesting us some corrections and
for pointing out to us ref.~\cite{HR}.
\end{acknowledgments}

\appendix*
\section{Derivation of the expressions for the time derivative of the entropy}
Let us first derive eq.~(\ref{dS2}). After substitution of
equations (\ref{kram}) and (\ref{op:lio}) into eq.~(\ref{dS1}), we
obtain
\begin{eqnarray}
\frac{\D S(t)}{\D t}&=&-\int \D x\; \ln
P\left\{\frac{\partial}{\partial\br }
\cdot\left(-\frac{\bp}{m}P\right)+\frac{\partial}{\partial\bp}\cdot
    \left[\left(\zeta \frac{\bp}{m}
      +\frac{\partial U_\mu}{\partial \br}
      -\bof_\mu\right)P+\zeta T \frac{\partial}{\partial\bp}
      P\right]\right\}\nonumber\\
&=& \int \D x\;\left\{ \de{\ln P}{\br}
\cdot\left(-\frac{\bp}{m}P\right) +\frac{\partial\ln
P}{\partial\bp}\cdot
    \left[\left(\zeta \frac{\bp}{m}
      +\frac{\partial U_\mu}{\partial \br}
      -\bof_\mu\right)P+\zeta T \frac{\partial}{\partial\bp}
      P\right]\right\}\nonumber\\
&=&\int \D x\; \left[P \zeta \frac{\bp}{ m }\cdot
\frac{\partial\ln P}{\partial\bp}
+\zeta T P \left(\frac{\partial\ln P}{\partial\bp}\right) ^2\right]\nonumber\\
&=&\int \D x\; \left[\zeta T \left(\frac{\bp}{T m}+ \de {\ln
P}{\bp}\right)^2  -\frac \zeta T \frac{\bp}{ m
}\cdot\left(\frac{\bp}{ m }+ \de {\ln P}{\bp}\right)
\right]P(x,t), \label{appdS1}
\end{eqnarray}
as can be checked by some manipulation. Equation~(\ref{appdS1})
corresponds thus to eq.~(\ref{dS2}).

We can now prove eq.~(\ref{ds1}). We have
\begin{equation}
\average{\left.\frac{\D s(x(t),t)}{\D
t}\right|_{x(t)=x}}_{\be}=\average{\left.\left[\de{s(x(t),t)}{\br}\cdot\dot
\br +\de{s(x(t),t)}{\bp}\cdot\dot
\bp+\de{s(x(t),t)}{t}\right]\right|_{x(t)=x}}_{\be}. \label{appds0}
\end{equation}
The last term on the rhs is independent of the path, and its
average vanishes because of the normalization. Thus we are left
with
\begin{eqnarray}
\average{\left.\frac{\D s(x(t),t)}{\D t}\right|_{x(t)=x}}_{\be}&=&
\int \D x\; P(x,t)\left\{\de{s(x,t)}{\br}\cdot \left[\dot
\br\right]_{\be} +\de{s(x,t)}{\bp}\cdot\left[\dot
\bp\right]_{\be}\right\}, \label{appds1}
\end{eqnarray}
where $\left[\cdots\right]_{\be}$ is the average over the process
realization alone, constrained by the final state $x(t)=x$. For
the process described by eqs.~(\ref{kram}) and (\ref{op:lio}), one
obtains
\begin{eqnarray}
\left[\dot \br\right]_{\be}&=&\frac {\bp}{m};\\
\left[\dot \bp\right]_{\be}&=&-\zeta \frac \bp m -\de
{U_{\mu}}{\br}+\bof_{\mu}-\zeta T\de{\ln P(x,t)}{\bp}.
\end{eqnarray}
Hence, eq.~(\ref{appds1}) becomes
\begin{eqnarray}
\average{\left.\frac{\D s(x(t),t)}{\D t}\right|_{x(t)=x}}_{\be}&=&
-\int \D x\; P \left\{\de{\ln P}{\br}\cdot\frac {\bp}{m}+\de{\ln
P}{\bp}\cdot\left[-\zeta \frac \bp m -\de
{U_{\mu}}{\br}+\bof_{\mu}-\zeta T\de{\ln P}{\bp}\right]\right\}
\nonumber\\
&=&\int \D x\;  P\,\de{\ln P}{\bp}\cdot\left[\zeta \frac \bp m
 +\zeta T\de{\ln P}{\bp}\right]\nonumber\\
&=&\int \D x\; P\,\left[\zeta T \left(\frac {\bp}{T m} + \de {\ln
P}{\bp}\right)^2  -\zeta  \frac {\bp}m\cdot \left(\frac {\bp} {Tm}
+ \de {\ln P}{\bp}\right) \right], \label{appds2}
\end{eqnarray}
which corresponds to eq.~(\ref{ds1}).

\end{document}